\documentclass[cup6b,epsf]{cupbook}
\usepackage{epsf}
\usepackage{psfig}
\usepackage{natbib}

\def\ojet{\Omega_{\rm jet}}

\def\eiso{E_{\rm iso}}
\def\egamma{E_{\gamma}}

\begin{document}

\author{D. Q. LAMB, T. Q. DONAGHY and C. GRAZIANI\\
Department of Astronomy \& Astrophysics, University of Chicago, \\
Chicago, IL 60637, USA}

\chapter{Gamma-Ray Bursts as a Laboratory for the Study of 
Type Ic Supernovae}

{\it 
HETE-2 has confirmed the connection between GRBs and Type Ic
supernovae.  Thus we now know that the progenitors of long GRBs are
massive stars.  HETE-2 has also provided strong evidence that the
properties of X-Ray Flashes (XRFs) and GRBs form a continuum, and
therefore that these two types of bursts are the same phenomenon.  We
show that both the structured jet and the uniform jet models can
explain the observed properties of GRBs reasonably well.  However, if
one tries to account for the properties of both XRFs and GRBs in a
unified picture, the uniform jet model works reasonably well while the
structured jet model fails utterly.  The uniform jet model of XRFs and
GRBs implies that most GRBs have very small jet opening angles ($\sim$
half a degree).  This suggests that magnetic fields play a crucial role
in GRB jets.  The model also implies that the energy radiated in gamma
rays is $\sim$ 100 times smaller than has been thought.  Most
importantly, the model implies that there are $\sim 10^4 -10^5$ more
bursts with very small jet opening angles for every such burst we see. 
Thus the rate of GRBs could be comparable to the rate of Type Ic core
collapse supernovae.  Accurate, rapid localizations of many XRFs,
leading to identification of their X-ray and optical afterglows and the
determination of their redshifts, will be required in order to confirm
or rule out these profound implications.  HETE-2 is ideally suited to
do this (it has localized 16 XRFs in $\sim$ 2 years), whereas {\it
Swift} is not.  The unique insights into the structure of GRB jets, the
rate of GRBs, and the nature of Type Ic supernovae that XRFs may
provide therefore constitute a compelling scientific case for
continuing HETE-2 during the {\it Swift} mission.
}

\section{Introduction}

Gamma-ray bursts (GRBs) are the most brilliant events in the Universe.
Long regarded as an exotic enigma, they have taken center stage in
high-energy astrophysics by virtue of the spectacular discoveries of
the past six years.  It is now clear that they also have important
applications in many other areas of astronomy: GRBs mark the moment of
``first light'' in the universe; they are tracers of the star
formation, re-ionization, and metallicity histories of the universe;
and they are laboratories for studying core-collapse supernovae.  It is
the last topic that we focus on here.

\section{GRB -- SN Connection}

There has been increasing circumstantial and tantalizing direct
evidence in the last few years that GRBs are associated with core
collapse supernovae [see, e.g. \cite{lamb2000}].  The detection and
localization of GRB 030329 by HETE-2 \citep{vanderspek2003a} led to a
dramatic confirmation of the GRB -- SN connection.  GRB 030329 was
among the brightest 1\% of GRBs ever seen (see Figure 2).  Its  optical
afterglow was $\sim 12^{\rm th}$ magnitude at 1.5 hours after the burst
\citep{price2003} -- more than 3 magnitudes brighter than the famous
optical afterglow of GRB 990123 at a similar time 
\citep{akerlof1999}.  In addition, the burst source and its
host galaxy lie very nearby, at a redshift $z = 0.167$
\citep{greiner2003}.  Given that GRBs typically occur at $z$ = 1-2, the
probability that the source of an observed burst should be as close as
GRB 030329 is one in several thousand.  It is therefore very unlikely
that HETE-2, or even {\it Swift}, will see another such event.

\begin{figure}[t]
\centerline{\hbox{
\psfig{file=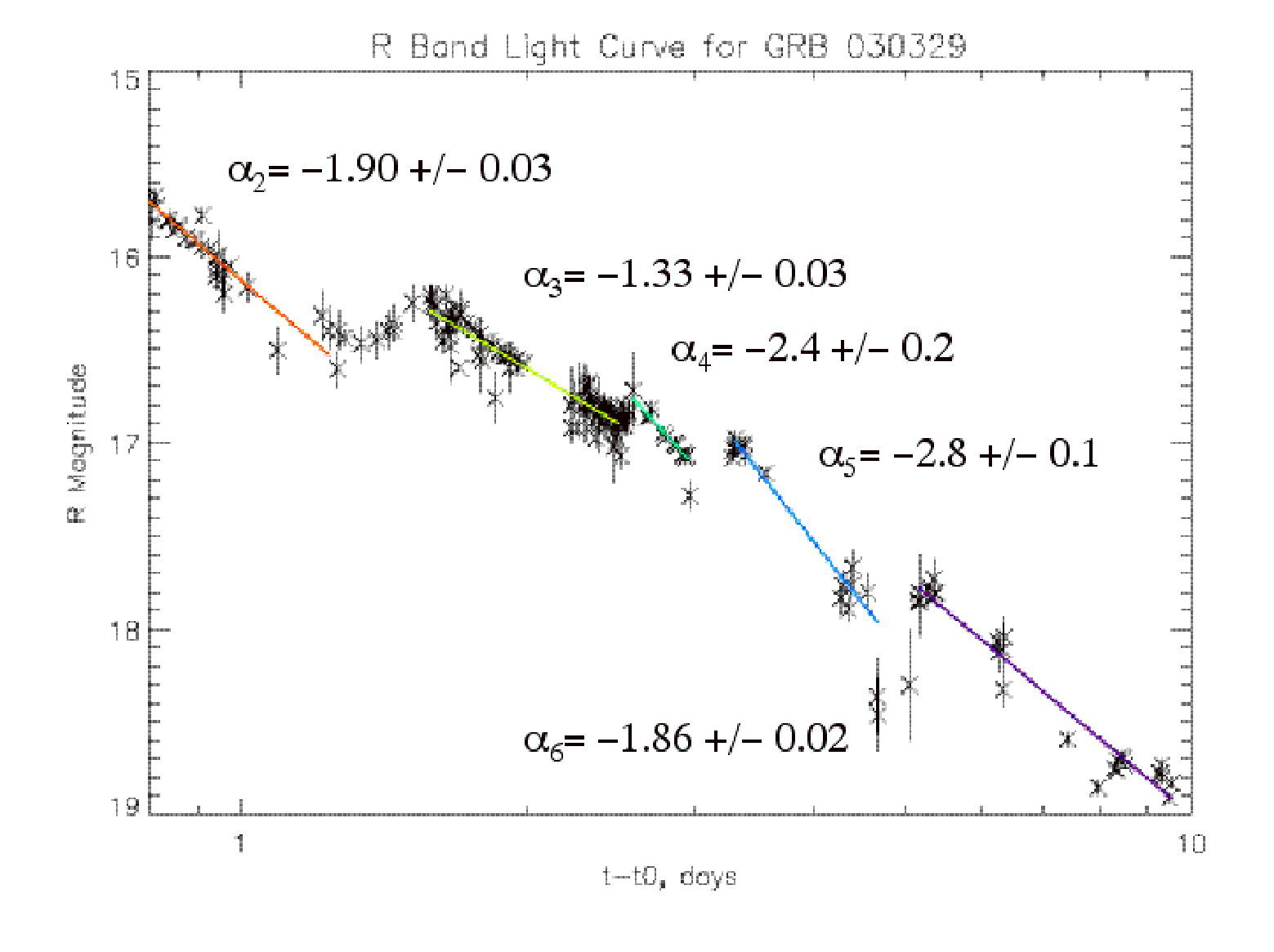,width=0.52 \textwidth}
\psfig{file=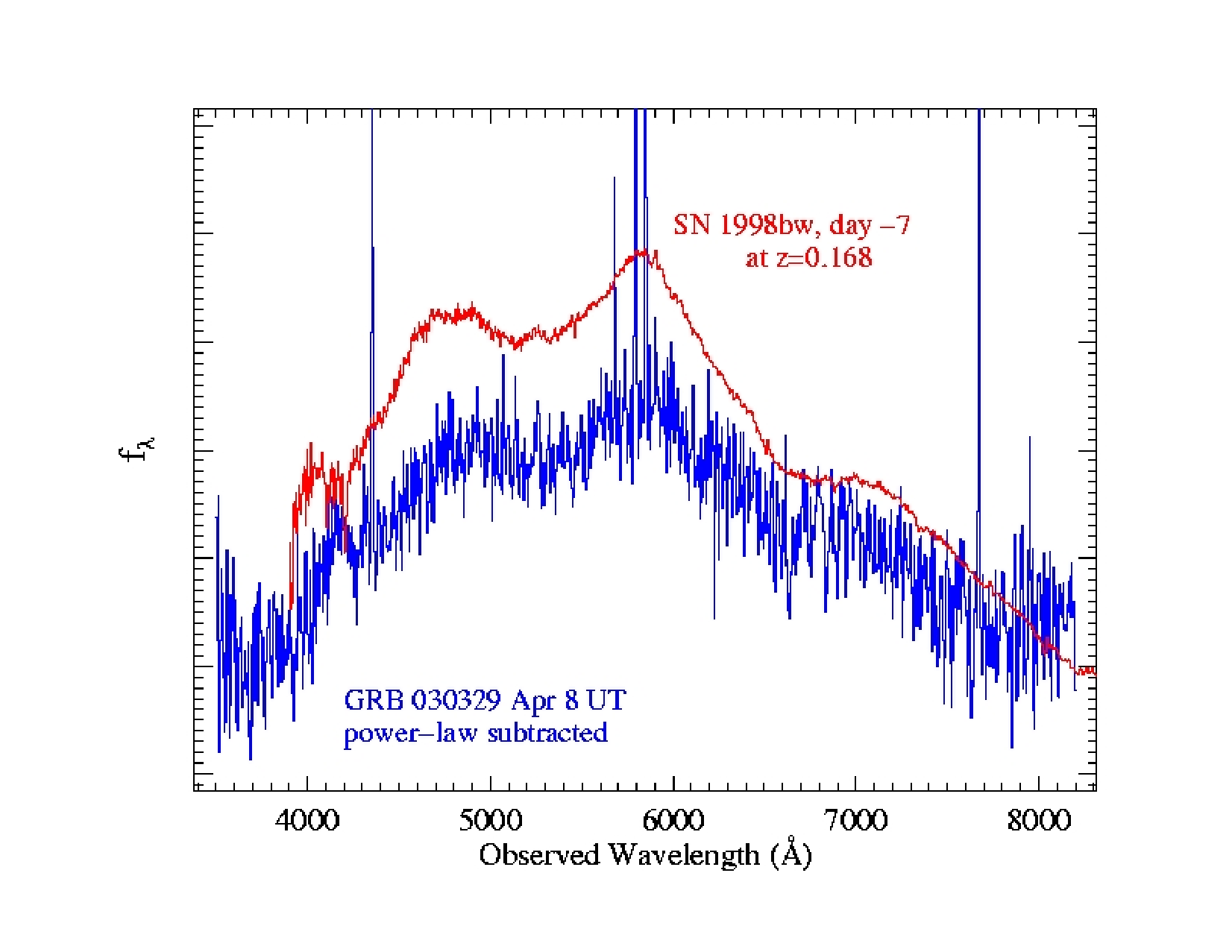,width=0.48 \textwidth}
}} 
\caption{Left panel:  Successive rebrightenings of the optical
afterglow of GRB 030329 during the 10 days following the burst.  From
\citep{filippenko2003}.  Right panel: Comparison of the discovery
spectrum of SN 2003dh seen in the  afterglow of GRB 030329 at 8 days
after the burst and the spectrum of the Type Ic supernova SN 1998bw. 
The similarity is striking.  From \cite{stanek2003}.
\label{fig3}}
\end{figure}

The fact that GRB 030329 was very bright spurred the astronomical
community -- both amateurs and professionals -- to make an
unprecedented number of observations of the optical afterglow of this
event.  Figure 1.1 (left panel) shows the light curve of the optical
afterglow of GRB 030329  1-10 days after the burst.  At least four
dramatic ``re-brightenings'' of the afterglow are evident in the
saw-toothed lightcurve.  These may be due to repeated injections of
energy into the GRB jet by the central engine at late times, or caused
by the ultra-relativistic jet ramming into dense blobs or shells of
material \citep{granot2003}.  If the former, it implies that the central
engine continued to pour out energy long after the GRB was over; if the
latter, it likely provides information about the last weeks and days of
the progenitor star.

The fact that GRB 030329 was very nearby made its optical afterglow an
ideal target for attempts to confirm the conjectured association
between GRBs and core collapse SNe.  Astronomers were not disappointed:
about ten days after the burst, the spectral signature of an energetic
Type Ic supernova emerged \citep{stanek2003}.  The supernova has been
designated SN 2003dh.  Figure 1.1 (right panel) compares the discovery
spectrum of SN 2003dh in the afterglow light curve of GRB 030329 and
the spectrum of the Type Ic supernova SN 1998bw.  The similarity is
striking.  The breadth and the shallowness of the absorption lines in
the spectra of SN 2003dh imply expansion velocities of $\approx$ 36,000
km s$^{-1}$ -- far higher than those seen in typical Type Ic
supernovae, and higher even than those seen in SN 1998bw.  It had been
conjectured that GRB 980425 was associated with SN 1998bw [see, e.g.,
\cite{galama1998}], but the fact that, if the  association were true,
the burst would have had to have been $\sim 10^4$ times fainter than
any other GRB observed to date made the association suspect.  The clear
detection of SN 2003dh in the afterglow of GRB 030329 confirmed
decisively the connection between GRBs and core collapse SNe.

The association between GRB 030329 and SN 2003dh makes it clear that we
must understand Type Ic SNe in order to understand GRBs.  The converse
is also true: we must understand GRBs in order to fully understand Type
Ic SNe.  It is possible that the creation of a powerful
ultra-relativistic jet as a result of the collapse of the core of a
massive star to a black hole plays a direct role in Type Ic supernova
explosions \citep{macfadyen2001}, but it is certain that the rapid
rotation of the collapsing core implied by such jets must be an
important factor in some -- perhaps most -- Type Ic supernovae.   The
result will often be a highly asymmetric explosion, whether the result
of rapid rotation alone or of the creation of powerful magnetic fields
as a result of the rapid rotation \citep{khokhlov1999}.

The large linear polarizations measured in several bright GRB
afterglows, and especially the temporal variations in the linear
polarization [see, e.g., \cite{rol2003}], provide strong evidence that
the Type Ic supernova explosions associated with GRBs are highly
asymmetric.  The recent dramatic discovery that GRB 021206 was strongly
polarized \citep{coburn2003} provides compelling evidence that GRB jets
are in fact dominated by magnetic energy rather than hydrodynamic
energy.

In addition, the X-ray afterglows of several GRBs have provided
tantalizing evidence of the presence of emission lines of
$\alpha$-particle nuclei \citep{reeves2002,butler2003}.  These
emission lines, if confirmed, provide severe constraints on models of
GRBs and Type Ic supernovae [see, e.g., \cite{lazzati2002}].  They may
also provide information on the abundances and properties of heavy
elements that have been freshly minted in the supernova explosion.

It is therefore now clear that GRBs are a unique laboratory for
studying, and are a powerful tool for understanding, Type Ic core
collapse supernovae.

\section{Nature of X-Ray Flashes and X-Ray-Rich GRBs}

Two-thirds of all HETE-2--localized bursts are either ``X-ray-rich'' or
X-Ray Flashes (XRFs); of these, one-third are XRFs \footnote{We define
``X-ray-rich'' GRBs and XRFs as those events for which $\log
[S_X(2-30~{\rm kev})/S_\gamma(30-400~{\rm kev})] > -0.5$ and 0.0,
respectively.} \citep{sakamoto2003b}.  These events have received
increasing attention in the past several years
\citep{heise2000,kippen2002}, but their nature remains unknown.

Clarifying the nature of XRFs and X-ray-rich GRBs, and their connection
to GRBs, could provide a breakthrough in our understanding of the
prompt emission of GRBs.  Analyzing 42 X-ray-rich GRBs and XRFs seen by
FREGATE and/or the WXM instruments on HETE-2, \cite{sakamoto2003b}
find that the XRFs, the X-ray-rich GRBs, and GRBs form a continuum in
the [$S_\gamma(2-400~{\rm kev}), E^{\rm obs}_{\rm peak}$]-plane (see
Figure 1.2, left-hand panel).  This result strongly suggests that all of
these events are the same phenomenon.

Furthermore, \cite{lamb2003c} have placed 9 HETE-2 GRBs with known
redshifts and 2 XRFs with known redshifts or strong redshift
constraints in the ($E_{\rm iso}, E_{\rm peak}$)-plane (see Figure 1.2,
right-hand panel).  Here $E_{\rm iso}$ is the isotropic-equivalent
burst energy and $E_{\rm peak}$ is the energy of the peak of the burst
spectrum, measured in the source frame.  The HETE-2 bursts confirm the
relation between $E_{\rm iso}$ and $E_{\rm peak}$ found by
\cite{amati2002} for GRBs and extend it down in $E_{\rm iso}$ by a
factor of 300.  The fact that XRF 020903, one of the softest events
localized by HETE-2 to date, and XRF 030723, the most recent XRF
localized by HETE-2, lie squarely on this relation
\citep{sakamoto2003a,lamb2003c} provides strong evidence that XRFs and
GRBs are the same phenomenon.  However, additional redshift
determinations are clearly needed for XRFs with 1 keV $< E_{\rm peak} <
30$ keV energy in order to confirm these results.

\begin{figure}[t]
\centerline{\hbox{
\psfig{file=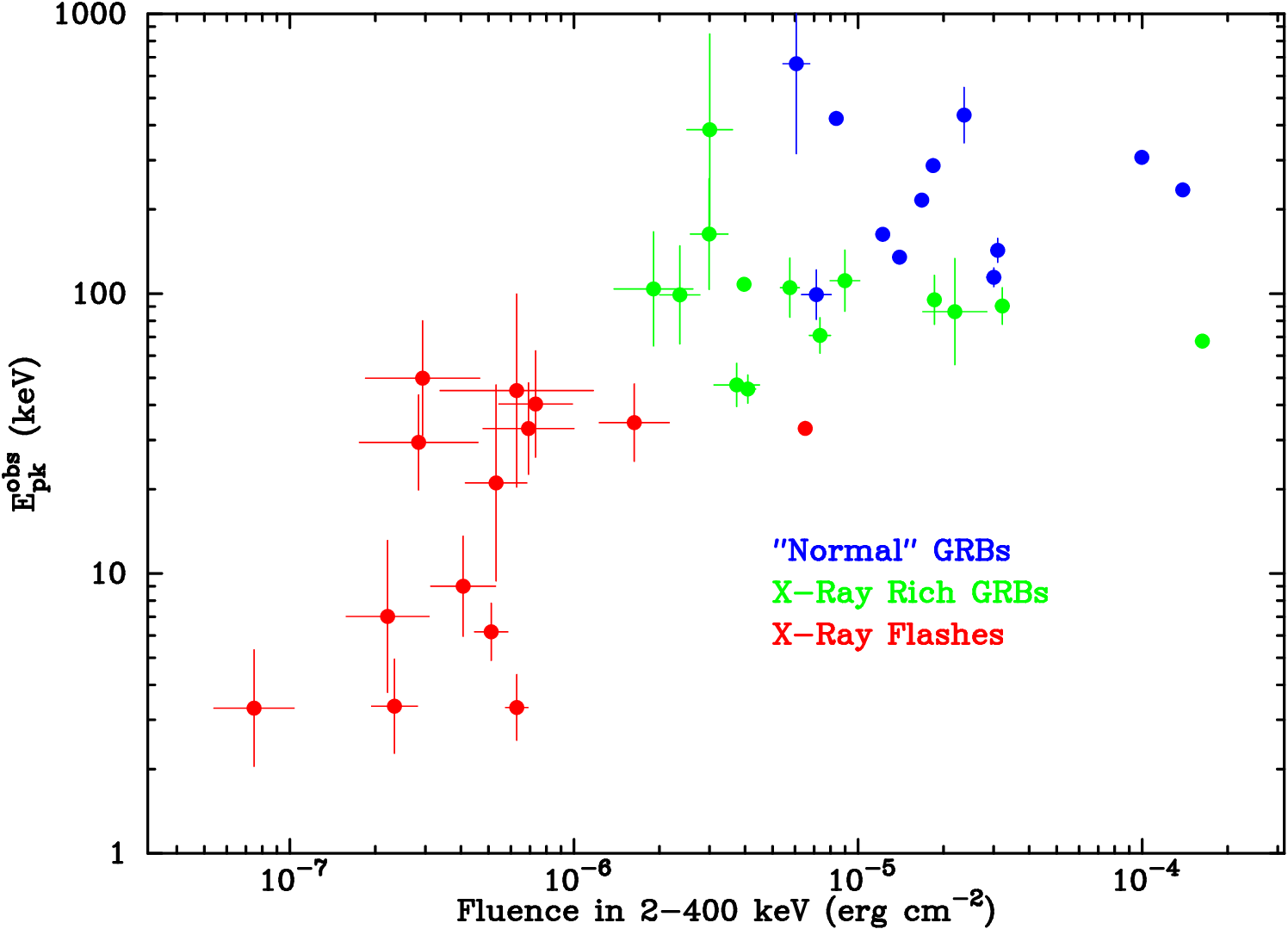,angle=270,width=0.554 \textwidth}
\psfig{file=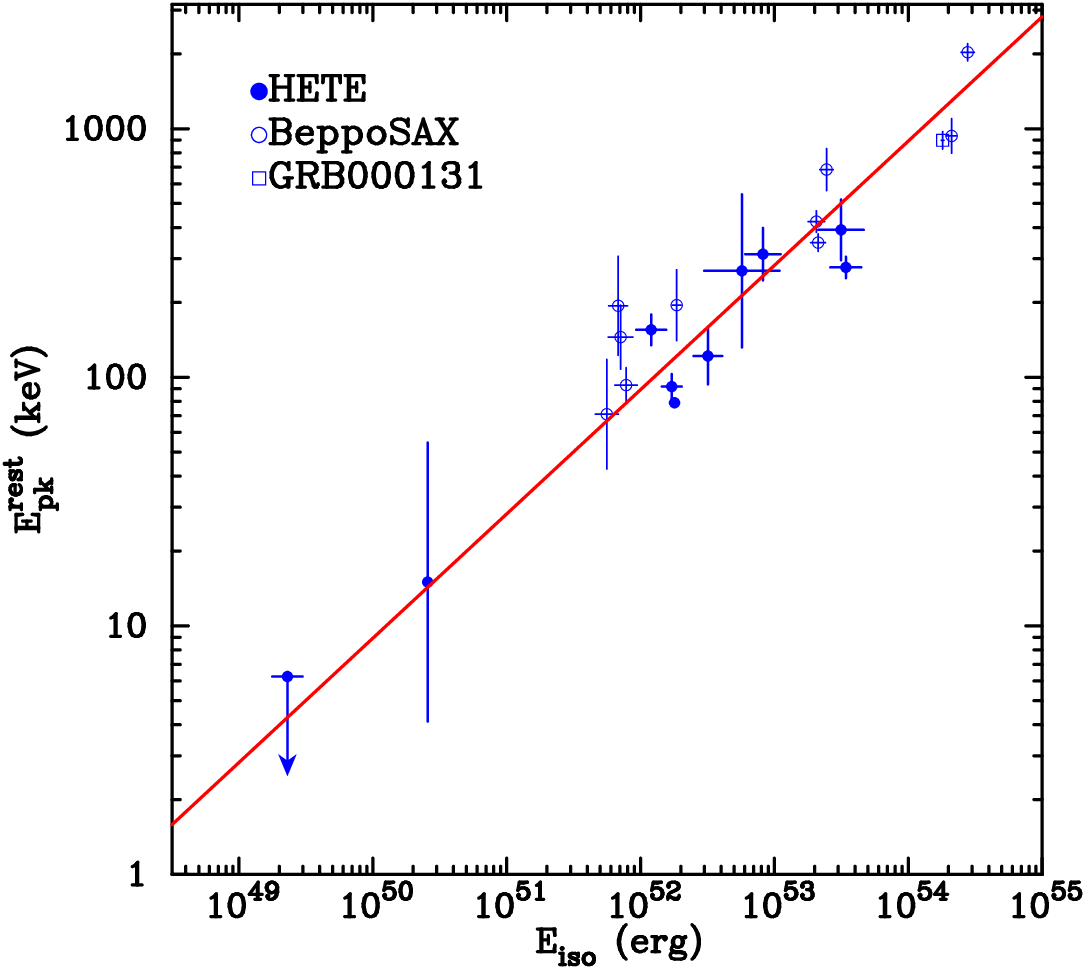,width=0.446 \textwidth}
}} 
\caption{Distribution of HETE-2 bursts in the [$S(2-400~{\rm keV}),
E^{\rm obs}_{\rm peak}$]-plane, showing XRFs (red), X-ray-rich GRBs
(green), and GRBs (blue) (left panel).    From \cite{sakamoto2003b}.
Distribution of HETE-2 and BeppoSAX bursts in the ($E_{\rm
iso}$,$E_{\rm peak}$)-plane, where $E_{\rm iso}$ and $E_{\rm peak}$ are
the isotropic-equivalent GRB energy and the peak of the GRB spectrum in
the source frame (right panel).   The HETE-2 bursts confirm the relation
between $E_{\rm iso}$ and $E_{\rm peak}$ found by Amati et al. (2002),
and extend it by a factor $\sim 300$ in $E_{\rm iso}$.  The bursts with
the lowest and second-lowest values of $E_{\rm iso}$ are XRFs 020903
and 030723. From \cite{lamb2003c}.
\label{fig16}}
\end{figure}
\vskip -0.2truein

\section{XRFs as a Probe of Type Ic Supernovae}

Frail et al. (2001; see also Bloom et al. 2003) have shown that most
GRBs have a ``standard'' energy; i.e, if their isotropic equivalent
energy is corrected for the jet opening angle inferred from the jet
break time, most GRBs have the same radiated energy, $\egamma = 1.3
\times 10^{51}$ ergs, to within a factor of $\pm$ 2-3.

\begin{figure}[t]
\centerline{
\psfig{file=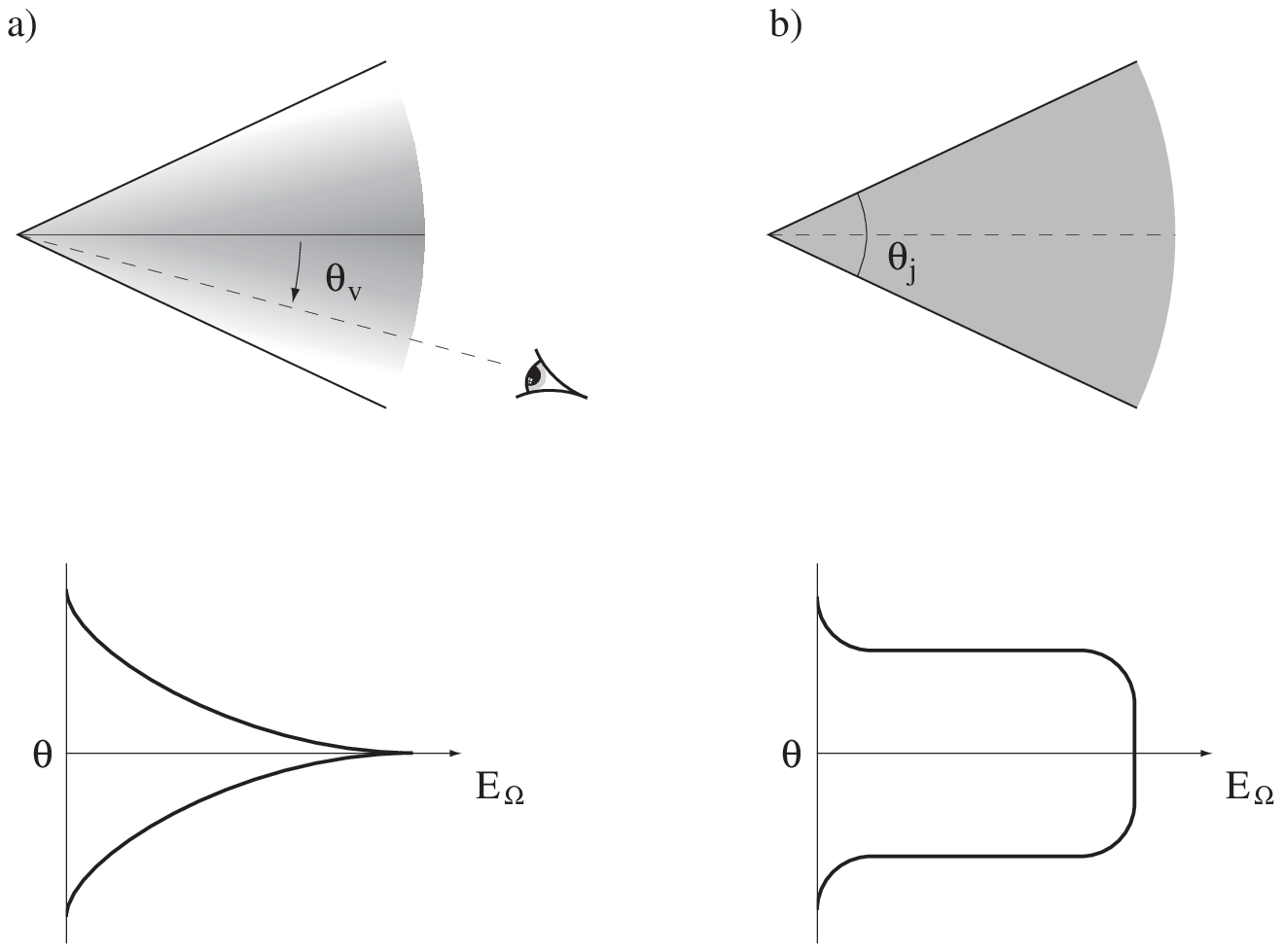,width=0.7 \textwidth}
}
\caption{
Schematic diagrams of universal jet model and jet model of GRBs
\citep{ramirez-ruiz2002}. 
In the universal jet model, the isotropic-equivalent energy and
luminosity is assumed to decrease as the viewing angle $\theta_v$ as
measured from the jet axis increases.  In order to recover the
``standard energy'' result \citep{frail2001}, $E_{\rm iso} (\theta_v) \sim
\theta_v^{-2}$ is required.  In the uniform jet model, GRBs produce
jets with a large range of jet opening angles  $\theta_{\rm jet}$.  For
$\theta < \theta_{\rm jet}$, $E_{\rm iso} (\theta_v)$ = constant while
for $\theta > \theta_{\rm jet}$, $E_{\rm iso} (\theta_v) = 0$.  
\label{fig20}}
\vskip -0.1truein
\end{figure}

Two models of GRB jets have received widespread attention:

\begin{itemize}

\item
The ``structured jet'' model (see the left-hand panel of Figure 1.3).
In this model, all GRBs produce jets with the same structure
\citep{rossi2002,woosley2003,zhang2002,meszaros2002}.  The
isotropic-equivalent energy and luminosity is assumed to decrease as
the viewing angle $\theta_v$ as measured from the jet axis increases. 
The wide range in values of $E_{\rm iso}$ is attributed  to differences
in the viewing angle $\theta_v$.  In order to recover the ``standard
energy'' result \citep{frail2001}, $E_{\rm iso} (\theta_v) \sim
\theta_v^{-2}$ is required \citep{zhang2002}.
\bigskip

\item
The ``uniform jet'' model (see the right-hand panel of Figure 1.3).  In
this model GRBs produce jets with very different jet opening angles 
$\theta_{\rm jet}$.  For $\theta < \theta_{\rm jet}$, $E_{\rm iso}
(\theta_v)$ = constant while for $\theta > \theta_{\rm jet}$, $E_{\rm
iso} (\theta_v) = 0$.

\end{itemize}

As we have seen, HETE-2 has provided strong evidence that the
properties of XRFs, X-ray-rich GRBs, and GRBs form a continuum, and
that these bursts are therefore the same phenomenon.  If this is true,
it immediately implies that the $\egamma$ inferred by 
\cite{frail2001} is too large by a factor of at least 100
\citep{lamb2003}.  The reason is that the values of $E_{\rm iso}$ for
XRF 020903 \citep{sakamoto2003a} and XRF 030723 \citep{lamb2003c} are
$\sim$ 100 times smaller than the value of  $\egamma$ inferred by Frail
et al. -- an impossibility.

HETE-2 has also provided strong evidence that, in going from XRFs to 
GRBs, $E_{\rm iso}$ changes by a factor $\sim 10^5$ (see Figure 1.2,
right-hand panel).  If one tries to explain only the range in $E_{\rm
iso}$ corresponding to GRBs, both the uniform jet model and the
structured jet model work reasonably well.  However, if one tries to
explain the range in $E_{\rm iso}$ of a factor $\sim 10^5$ that is
required in order to accommodate both XRFs and GRBs in a unified
description, the uniform jet works reasonably well while the structured
jet model fails utterly.

The reason is the following:  the observational implications of the
structured jet model and the uniform jet model differ dramatically if
they are required to explain XRFs and GRBs in a unified picture.  In
the structured jet model, most viewing angles $\theta_v$ are $\approx
90^\circ$.  This implies that the number of XRFs should exceed the
number of GRBs by many orders of magnitude, something that HETE-2 does
not observe (see Figures 1.2, 1.4, and 1.5).  On the other hand, by
choosing $N(\Omega_{\rm jet}) \sim \Omega_{\rm jet}^{-2}$, the uniform
jet model predicts equal numbers of bursts per logarithmic decade in
$E_{\rm iso}$ (and $S_E$), which is exactly what HETE-2 sees (again,
see Figures 1.2, 1.4, and 1.5) \citep{lamb2003}.

Thus, if $E_{\rm iso}$ spans a range $\sim 10^5$, as the HETE-2 results
strongly suggest, the uniform jet model can provide a unified picture
of both XRFs and GRBs, whereas the structured jet model cannot.  This
means that XRFs provide a powerful probe of GRB jet structure.

\begin{figure}[t]
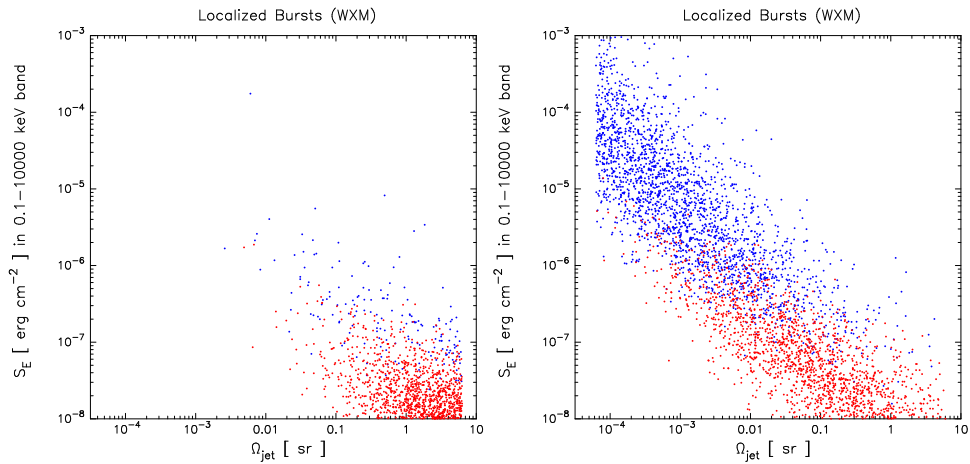

\centerline{\hbox{
\psfig{file=figures/omega_SE.univ1m.ps,angle=270,width=0.5 \textwidth}
\psfig{file=figures/omega_SE.m2.ps,angle=270,width=0.5 \textwidth}
}}
\caption{Expected distribution of bursts in the ($\Omega_{\rm
jet},S_E$)-plane for the universal jet model (left panel) and uniform
jet model (right panel), assuming that the Amati et al. (2002) relation
holds for XRFs as well as for GRBs, as the HETE-2 results strongly
suggest.  From \cite{lamb2003}.
\label{fig21}}
\end{figure}

A range in $\eiso$ of $10^5$, which is what the HETE-2 results strongly
suggest, requires a {\it minimum} range in $\Delta \ojet$ of $10^4 -
10^5$ in the uniform jet model.  Thus the unified picture of XRFs and
GRBs in the uniform jet model implies that there are $\sim
10^4 - 10^5$ more bursts with very small $\ojet$'s for every such burst
we see; i.e., the rate of GRBs may be $\sim 100$ times greater
than has been thought.

In addition, since the observed ratio of the rate of Type Ic
supernovae to the rate of GRBs in the observable universe is $R_{\rm
Type\ Ic}/ R_{\rm GRB} \sim 10^5$ \citep{lamb1999}, a unified picture of
XRFs and GRBs in the uniform jet model implies that roughly {\it
all} Type Ic supernovae produce high-energy transients
\citep{lamb2003}.  More spherically symmetric jets yield XRFs and narrow
jets produce GRBs.  Thus XRFs and GRBs provide a combination of GRB/SN
samples that would enable astronomers to study the relationship between
the degree of jet-like behavior of the GRB and the properties of the
supernova (brightness, polarization $\Leftrightarrow$ asphericity of
the explosion, velocity of the explosion $\Leftrightarrow$ kinetic
energy of the explosion, etc.).  GRBs may therefore provide a unique
laboratory for understanding Type Ic core collapse supernovae.

\begin{figure}[t]
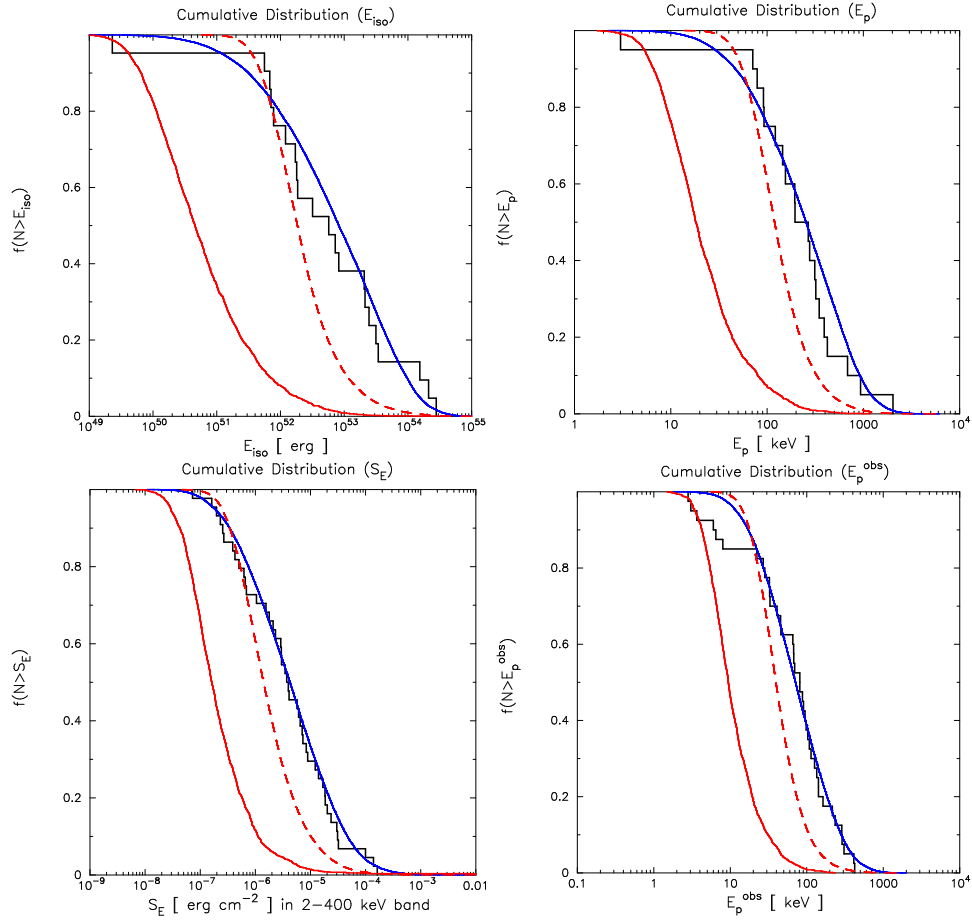

\centerline{\hbox{
\psfig{file=figures/both.Eiso.comp.cuml.ps,angle=270,width=0.5 \textwidth}
\psfig{file=figures/both.Ep.comp.cuml.ps,angle=270,width=0.5 \textwidth}
}}
\centerline{\hbox{
\psfig{file=figures/hete.flu.2_400.comp.cuml.ps,angle=270,width=0.5 \textwidth}
\psfig{file=figures/hete.Ep_obs.comp.cuml.ps,angle=270,width=0.5 \textwidth}
}}
\caption{Top row: cumulative distributions of $S(2-400 {\rm keV})$ (left panel)
and $E^{\rm obs}_{\rm peak}$ (right panel) predicted by the structured
(red) and uniform (blue) jet models, compared to the observed
cumulative distributions of these quantities.  Bottom row: cumulative
distributions of $E_{\rm iso}$ (left panel) and $E_{\rm peak}$ (right
panel) predicted by the structured (red) and uniform (blue) jet models,
compared to the observed cumulative distributions of these quantities. 
The cumulative distributions corresponding to the best-fit structured
jet model that explains XRFs and GRBs are shown as solid lines; the
cumulative distributions corresponding to the best-fit structured jet
model that explains GRBs alone are shown as dashed lines.  The
structured jet model provides a reasonable fit to GRBs alone but cannot
provide a unified picture of both XRFs and GRBs, whereas the uniform
jet model can.  From \cite{lamb2003}.
\label{fig22}}
\end{figure}

A unified picture of XRFs and GRBs in the uniform jet model also
implies that most Type Ic supernovae produce narrow jets, which may
suggest that the collapsing cores of most Type Ic supernovae are
rapidly rotating.  Finally, such a unified picture implies that the
total radiated energy in gamma rays $\egamma$ is $\sim 100$ times
smaller than has been thought \citep{lamb2003}.  

As we have seen, the HETE-2 results provide strong evidence that XRFs
and GRBs are the same phenomenon.  But the profound implications of
these results in terms of the structure of GRB jets, the rate of GRBs,
and the nature of Type Ic supernovae, require incontrovertible
evidence.  

Obtaining the incontrovertible evidence needed to sustain (or refute) 
these implications will require accurate, rapid localizations of XRFs,
leading to identification of their X-ray and optical afterglows and the
determination of their redshifts.  Until very recently, only one XRF
(XRF 020903; Soderberg et al. 2002) had even a probable optical
afterglow and redshift.  The reason why is that, as expected in the
uniform jet picture, the X-ray (and therefore the optical) afterglows
of XRFs are $\sim 10^3 - 10^4$ times fainter than those of GRBs
\citep{lamb2003}.  But this challenge can be met: the recent
HETE-2--localization of XRF 030723 represents the first time that an
XRF has been localized in real time \citep{prigozhin2003};
identification of its X-ray and optical afterglows rapidly followed
\citep{fox2003c}.  This event may well be the Rosetta stone for XRFs.

The exciting recent results involving XRF 030723 highlight the fact
that HETE-2 is ideally suited to obtain the evidence about XRFs that is
required to confirm or rule out the profound implications about the
structure of GRB jets, the rate of GRBs, and the nature of Type Ic
supernovae described above.  HETE-2 will obtain this evidence, if the
HETE-2 mission is extended, whereas {\it Swift} cannot.  HETE-2's
ability to accurately and rapidly localize XRFs -- and  study their
spectra -- therefore constitutes a compelling reason for continuing
HETE-2 during the {\it Swift} mission.

\end{document}